\documentclass[]{aa}
\usepackage[varg]{txfonts}
\usepackage{graphicx}
\usepackage{subfig}
\usepackage{stackengine}
\usepackage[pscoord]{eso-pic}
\usepackage[export]{adjustbox}
\usepackage{color}
\bibpunct{(}{)}{;}{a}{}{,} 

\title{Evolution of the Sun's non-axisymmetric toroidal field}
\author{D. Martin-Belda\inst{\ref{inst1}, \ref{inst2}}
\and R. H. Cameron\inst{\ref{inst2}}
}

\institute{Institut für Astrophysik, Georg-August-Universität Göttingen, 37077 Göttingen, Germany\label{inst1} \and Max-Planck-Institut für Sonnensystemforschung, Justus-von-Liebig-Weg 3, 37077 Göttingen, Germany\label{inst2}
}

\date{Received / Accepted}


\abstract
{}
{We aim to infer the sub-surface distribution of the Sun's non-axisymmetric azimuthal magnetic flux from observable quantities, such as the surface magnetic field and the large scale plasma flows.}
{We have built a kinematic flux transport model of the solar dynamo based on the Babcock-Leighton framework. We constructed the source term for the poloidal field using SOLIS magnetograms spanning three solar cycles. Based on this source we calculated the azimuthal flux below the surface. The flux transport model has two free parameters which we constrain using sunspot observations from cycle 22. We compared the model results with observations from cycle 23.}
{The structure of the azimuthal field is mainly axisymmetric. The departures from axisymmetry represent, on average, $\sim3\%$ of the total azimuthal flux. Owing to its relative weakness, the non-axisymmetric structure of the azimuthal field does not have a significant impact on the location in which the emergences appear or on the amount of flux contained in them. We find that the probability of emergence is a function of the ratio between the flux content of an active region and the underlying azimuthal flux.}
{}

\begin{document}
\titlerunning{Toroidal field}
\authorrunning{D. Martin-Belda \& R. H. Cameron}
\maketitle

\section{Introduction}\label{sec:intro}

The magnetic activity of the Sun and other stars is a manifestation of their internal magnetic field, which is thought to be sustained by a hydromagnetic dynamo. In the case of the Sun, it is generally thought that the differential rotation in the convection zone generates the toroidal magnetic field out of the poloidal field, but where exactly this field is amplified and stored is still an open question \citep[see, e.g.][]{charbonneau2010lrsp}. The mechanism for the regeneration of the poloidal field from the azimuthal component is less agreed upon, with modelling approaches falling mainly into two categories: the turbulent dynamo models and the Babcock-Leighton models. 

In Babcock-Leighton models \citep{babcock1961topology,leighton1969kinematic}, the poloidal field is regenerated by the surface transport of the magnetic flux of decaying active regions. Newly emerged bipolar magnetic regions (BMRs) show a systematic tilt with respect to the E-W direction, with the preceding polarity (in the Sun's sense of rotation) appearing closer to the equator than the trailing polarity (Joy's law). In addition, the preceding polarity of a BMR emerging in a given hemisphere tends to be of the same sign as the polar field in that hemisphere at the beginning of the ongoing activity cycle (Hale's law). This facilitates the cross-equatorial transport of preceding polarity flux, and leads to the cancellation of the polar fields and the build-up of a new, reversed axial dipole, which is the source of azimuthal field for the new activity cycle.

Babcock-Leighton models have gained substantial support in recent years. \cite{dasi2010angles} found a strong correlation between the strength of a Babcock-Leighton type source term in a given cycle, calculated from the observed tilt angle of active regions, and the strength of the next  cycle. \cite{kitchatinov2011babcock} found that the aggregate contribution of active regions to the poloidal field during one cycle and the strength of the global dipole at the end of the same cycle (as inferred from the AA index) also correlate closely. \cite{wang2009polar} showed that the build-up of the polar fields during cycles 20 to 23 is consistent with the passive transport of magnetic flux by the observed surface flows. On the theoretical side, \cite{cameron2015crucial} showed that the main source of net azimuthal flux in each hemisphere is the winding up of poloidal flux that is connected to the polar fields at the surface.

\begin{figure*}[t]
  \centering
    \subfloat{
      \centering
      \includegraphics[width=.47\textwidth]{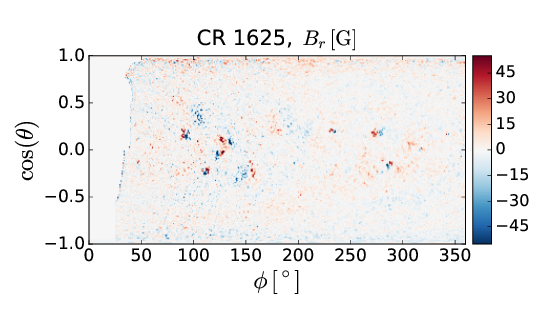}
    }
    \subfloat{
      \centering
      \includegraphics[width=.47\textwidth]{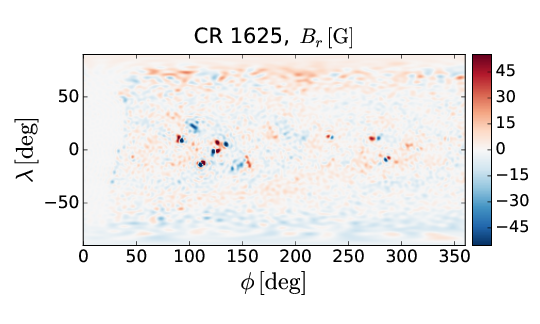}
    }\\
    \subfloat{
      \centering
      \includegraphics[width=.47\textwidth]{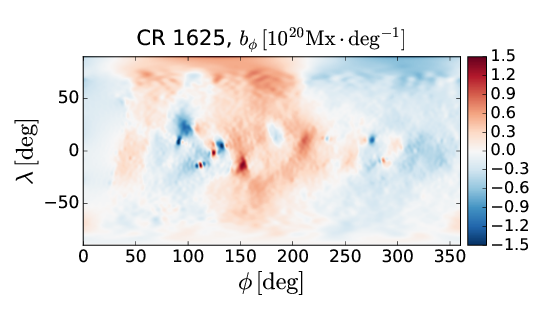}
    }
    \subfloat{
      \centering
      \includegraphics[width=.47\textwidth]{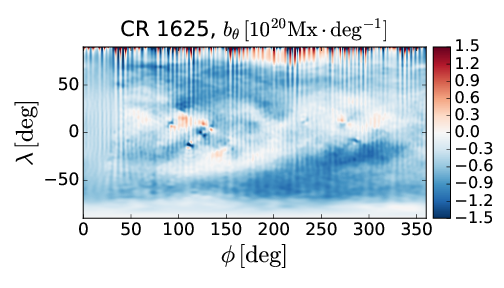}
    }
  \caption{Initial condition of the simulations. \textit{Top left:} Synoptic magnetogram corresponding to CR 1625. \textit{Top right:} Synoptic magnetogram corresponding to CR 1625, remapped to an equiangular grid and resampled to the highest angular degree order, $l$, used to compute the potential field extrapolation. \textit{Bottom left:} Radial integral of the azimuthal field across the convection zone, $b_\phi$, extrapolated from the surface field. \textit{Bottom right:} Radial integral of the $\theta$ component of the magnetic field across the convection zone, $b_\theta$, extrapolated from the surface field. Red and blue indicate opposite polarities in all the maps.}
  \label{fig:initCond}
\end{figure*}

One of the key features of dynamo models is the formulation of the poloidal source term as a function of the azimuthal field. One often considered possibility in Babcock-Leighton flux transport models is that magnetic flux tubes are stored in an overshoot region at the base of the convection zone. These develop a magnetic buoyancy instability and rise through the convection zone to emerge at the surface in the form of BMRs. For a review on this topic, see, for example, \cite{fan2009fields}.

Alternatively, 3D numerical calculations indicate that persistent, coherent azimuthal magnetic structures can arise in a turbulent convection zone, owing to turbulent intermittency \citep{brown2010wreaths, nelson2013wreaths, nelson2014buoyant}. Moreover, the transport of magnetic flux to the surface can be achieved by means of convective upflows, which might be at least as relevant as magnetic buoyancy.

In this work we consider the evolution of the radially integrated azimuthal flux density as a function of longitude and latitude. Our aim is to infer the distribution and evolution of the sub-surface azimuthal flux from observable quantities in order to gain insight on its relation with the observed properties of active regions. To do so, we have constructed a model of the magnetic flux transport in the Sun, based on the Babcock-Leighton framework. The poloidal field source term is determined by observational data (synoptic magnetograms). The paper is structured as follows: in Sect. \ref{sec:methods} we introduce and calibrate our model; in Sect. \ref{sec:results} we present and discuss our results; and in Sect. \ref{sec:conclusion} we briefly summarize our conclusions. In the appendices we derive the equation for the evolution of the azimuthal flux density and other supplementary quantities in our model.

\section{Methods}\label{sec:methods}

\subsection{Model}\label{sec:model}

We have considered a mean field approach based on horizontal averages,
\begin{equation}
B_j = \langle B_j \rangle +B'_j,
\end{equation}
where $j=\{r,\theta,\phi\}$, 
\begin{equation}
\langle B_j \rangle (\theta,\phi) = \frac{\int_{\theta-\delta}^{\theta+\delta}\int_{\phi-\delta}^{\phi+\delta}B_j\sin\theta'\,\mathrm{d}\phi'\mathrm{d}\theta'}{\int_{\theta-\delta}^{\theta+\delta}\int_{\phi-\delta}^{\phi+\delta}\sin\theta'\,\mathrm{d}\phi'\mathrm{d}\theta'},
\label{eq:meanField}
\end{equation}
and $\delta$ is the scale over which the average is performed. Using $\delta\sim2-3^\circ$ is enough to ensure some scale separation with respect to the larger turbulence correlation lengths at the surface (those of supergranulation). In the remainder of the paper we drop the angle brackets for clarity, and refer to the $j$-th component of the averaged magnetic field by $B_j$.

Our model consists of two two-dimensional domains in the $(\phi,\theta)$ plane, representing the surface of the Sun and the convection zone, respectively. The evolution of the surface magnetic field, assumed to be radial, is governed by the surface flux transport equation \citep{devore1984meridional}:
\begin{align}
    \frac{\partial B_r}{\partial t} = &-\Omega_{R_\odot}(\theta)\frac{\partial B_r}{\partial\phi}-\frac{1}{R_\odot \sin\theta}\frac{\partial}{\partial\theta}\left[ u_M(\theta)B_r\sin\theta\right] \nonumber\\ 
    &+ \frac{\eta_H}{R_\odot^2}\left[\frac{1}{\sin\theta}\frac{\partial}{\partial\theta}\left(\sin\theta\frac{\partial B_r}{\partial\theta}\right)+\frac{1}{\sin^2\theta}\frac{\partial^2 B_r}{\partial\phi^2}\right] \nonumber\\
    &+S(\theta, \phi, t),
    \label{eq:sft}
\end{align}
where $\Omega_{R_\odot}$ is the differential rotation, $u_M$ is the meridional flow and $\eta_H$ is the surface diffusivity associated to the convective flows. The emergence of new flux is described by the source term $S(\theta, \phi, t)$.

The first term in Eq. \eqref{eq:sft} describes the transport of the surface field by the solar differential rotation. We use the differential rotation profile inferred by correlation tracking of magnetic features by \cite{hathaway2011variations}:
\begin{equation}
    \Omega_{R_\odot}(\theta) = 14.437-1.48\cos^2\theta-2.99\cos^4\theta\,\,\,\,\,\mathrm{[^\circ/day]}.
    \label{eq:diffRot}
\end{equation}

The second term in Eq. \eqref{eq:sft} corresponds to the surface meridional flow. Following \cite{vanBallegooijen1998fchannels}, we model the meridional flow as:
\begin{equation}
    u_M = -11\cos(2\theta)\,\,\,\,\,\mathrm{[ms^{-1}]}.
    \label{eq:mFlow}
\end{equation}
This expression captures the main characteristics of the observed meridional flow \citep[see][]{hathaway2011variations}.
 
The third term in Eq. \eqref{eq:sft} describes the dispersal of magnetic flux on the surface by means of random convective flows modelled as a diffusion process \citep[see][]{leighton1964transport,mbelda2015inflows}. We use a surface diffusivity of $\eta_H = 250\,\mathrm{km^2s^{-1}}$, as indicated by observations \citep{schrijver1990patterns,jafarzadeh2014migration}.

The source term $S(\theta,\phi,t)$ represents new emergences, and is built from synoptic magnetograms (see Sect. \ref{sec:sources}).

The azimuthal field in the convection zone is represented in our model by the azimuthal flux per unit colatitude, that is:
\begin{equation}
b_\phi = \int_{R_{\rm b}}^{R_\odot} B_\phi r \mathrm{d}r.
\end{equation}
In the above expression, $R_{\rm b}$ refers to the bottom of the convection zone and $R_\odot$ is the solar radius.

Following \cite{cameron2016update}, we made the following assumptions regarding the plasma flows and the structure of the internal magnetic field of the Sun:
\begin{enumerate}
\item The magnetic field is purely radial in the near-surface shear layer (NSSL), owing to strong downwards turbulent pumping. 
\item The magnetic field does not penetrate the radiative interior.
\item The poloidal field does not penetrate the tachocline. This assumption of the model is partly justified by \cite{spruit2011critical}, who noted that the tachocline cannot support large shear stresses, which would be present if the poloidal field did penetrate the tachocline. 
\item The radial shear is negligible in the region between the tachocline and the NSSL. This is based on helioseismic inference of the rotation rate in the deep interior and the NSSL \citep{christensen1988differential}.
\end{enumerate}
To derive an evolution equation for $b_\phi$, we integrated the azimuthal component of the induction equation in radius. The resulting equation reads (see Appendix \ref{ap:a}):
\begin{align}
    \frac{\partial b_\phi}{\partial t} & = R_\odot^2\,\Omega_{R_\odot} \sin\theta\, B_r \nonumber\\
    & + b_\theta\sin\theta \left.\frac{\mathrm{d}\Omega}{\mathrm{d}\theta}\right|_{R_{\rm NSSL}} + \frac{\partial (b_\theta\sin\theta)}{\partial\theta}\Omega_{R_{\rm NSSL}} \nonumber\\
    & + \frac{\eta_0}{R_\odot^2}\left[\frac{1}{\sin\theta}\frac{\partial}{\partial\theta}\left(\sin\theta\frac{\partial b_\phi}{\partial \theta} \right) + 2\frac{\cos\theta}{\sin^2\theta}\frac{\partial b_\theta}{\partial\phi}-\frac{b_\phi}{\sin^2\theta} \right. \nonumber\\
    & + \left. \frac{1}{\sin^2\theta}\frac{\partial^2 b_\phi}{\partial \phi^2} \right] - \frac{1}{R_\odot}\frac{\partial ( \bar{u} b_\phi)}{\partial\theta} + S_\phi(\theta,\phi,t),
    \label{eq:azimuthal}
\end{align}
where $\Omega_{R_{\rm NSSL}}$ is the differential rotation at the bottom of the NSSL, $\bar{u}$ is an effective return meridional flow, $\eta_0$ is the effective diffusivity of the azimuthal field, $S_\phi(\theta,\phi,t)$ is a source term associated to flux emergence, and 
\begin{equation}
b_\theta = \int_{R_{\rm {\rm T}}}^{R_{\rm NSSL}}B_\theta r\,\mathrm{d}r.
\label{eq:bThetaDef1}
\end{equation}
In the above expression, $R_{\rm T}$ refers to the top of the tachocline and $R_{\rm NSSL}$ refers to the bottom of the NSSL. The quantity $b_\theta$ is calculated in terms of $b_\phi$ and $B_r$ in Appendix \ref{ap:b}.

The first three terms in Eq. \eqref{eq:azimuthal} describe the generation of azimuthal flux by differential rotation and the azimuthal flux transport. Here, $B_r$ is the radial field at the surface, which is constrained by observation and evolves according to Eq. \eqref{eq:sft}. The radially integrated $\theta$-component of the magnetic field, $b_\theta$, can be obtained in terms of $b_\phi$ and $B_r$ from the solenoidality condition $\nabla\cdot\mathbf{B}=\mathbf{0}$ (see Appendix \ref{ap:b}). The differential rotation profile is evaluated at the bottom of the NSSL. The analysis of helioseismic data by \cite{barekat2014radial} suggests that the radial shear in this layer is independent of latitude. Following these authors, we adopted
\begin{equation}
\Omega_{R_{\rm NSSL}}(\theta) = \Omega_{R_\odot}(\theta)+0.53\,\,\,\,\,\mathrm{[^\circ/day]}.
\end{equation}

The fourth term of Eq. \eqref{eq:azimuthal} describes the turbulent diffusion of azimuthal flux. Following \cite{cameron2016diffusion}, we assumed the following form for the diffusivity in the derivation of Eq. \eqref{eq:azimuthal}:
\begin{equation}
    \eta(r) = \eta_0 \frac{r^2}{R_\odot^2},
    \label{eq:diffusivity}
\end{equation}
where $\eta_0$ is a free parameter of our model. \cite{cameron2016diffusion} used the properties of the decay phase of the sunspot cycles to estimate $\eta_0\sim150-450\,\mathrm{km^2s^{-1}}$.

\begin{figure}[t!]
  \centering
  \resizebox{\hsize}{!}{\includegraphics[width=\textwidth]{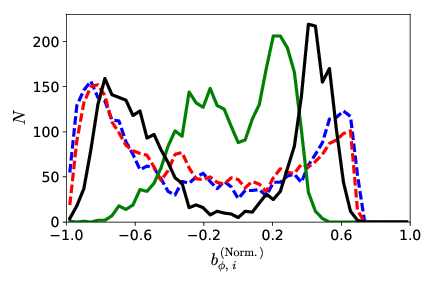}}
  \caption{Number of emergences ($N$) as a function of the normalized underlying azimuthal flux density ($b_{\phi,\,i}^{\rm (Norm.)}$). The different curves correspond to different choices of the free parameters. \textit{Continuous black line:} $u_0=3\,\mathrm{m\,s^{-1}}$ and $\eta_0=100\,\mathrm{km^2 s^{-1}}$. \textit{Dashed blue line:} $u_0=5\,\mathrm{m\,s^{-1}}$ and $\eta_0=400\,\mathrm{km^2 s^{-1}}$. \textit{Dashed red line:} $u_0=1\,\mathrm{m\,s^{-1}}$ and $\eta_0=25\,\mathrm{km^2 s^{-1}}$. \textit{Continuous green line:} $u_0=6\,\mathrm{m\,s^{-1}}$ and $\eta_0=600\,\mathrm{km^2 s^{-1}}$.}
  \label{fig:calibration}
\end{figure}

\begin{figure}[t]
  \centering
  \resizebox{\hsize}{!}{\includegraphics[width=\textwidth]{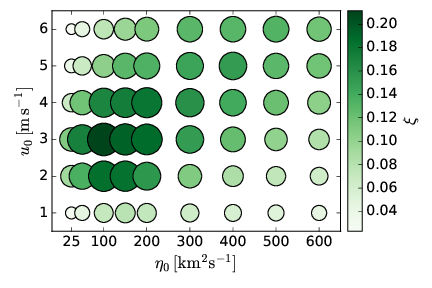}}
  \caption{Distribution of the quantity $\left. \xi = \sum_c N_c |b_{\phi,\,c}|  \middle/ \sum_c N_c / |b_{\phi,\,c}|\right.$ as a function of the parameters $u_0$ and $\eta_0$. Each circle represents a run. The size of each circle is proportional to the quantity $\xi$, which measures the adjustment of the run to our requirement that the simulated azimuthal flux lies underneath the observed active regions. For better visualization, we also encode the value of $\xi$ in the colour of the circles.}
  \label{fig:calibration2}
\end{figure}

The fifth term corresponds to the advection of the azimuthal flux by an effective equatorward flow, which we modeled as
\begin{equation}
    \bar{u} = u_0 \cos(2\theta)\,\,\,\,\,\mathrm{[ms^{-1}]},
    \label{eq:mFlow2}
\end{equation}
where $u_0$ is a free parameter of the model. This flow can correspond to a return meridional flow \citep{wang1991model, durney1995model} or equatorward pumping \citep{guerrero2008pumping}.

The solenoidality condition requires that the surface magnetic field connects to the sub-surface field. Hence, a source term in Eq. \eqref{eq:azimuthal} is needed to ensure the connectivity of the surface field sources to the field in the convection zone. In order to calculate $S_\phi(\theta,\phi,t)$, we extrapolated the surface sources downward via a potential field solution (see Sect. \ref{sec:sources}).

Our model, therefore, consists of: (a) a two-dimensional domain representing the surface of the Sun, in which the (radial) surface field evolves according to Eq. \eqref{eq:sft}; (b) a two-dimensional domain representing the convection zone, in which the radial integral of the azimuthal magnetic field evolves according to Eq. \eqref{eq:azimuthal}; and (c) the coupling of both domains through the emergences, represented by the source terms, and the solenoidality condition.

\subsection{Treatment of the source terms}\label{sec:sources}

\begin{figure*}[t]
  \centering

    \subfloat{
      \centering
      \includegraphics[width=.47\textwidth]{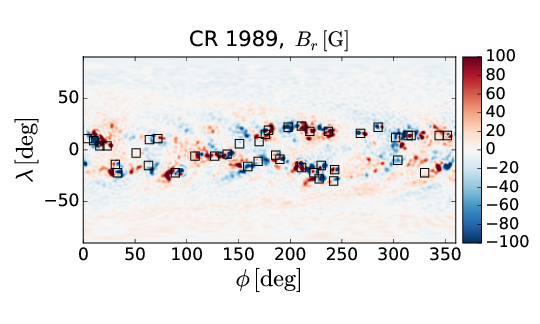}
    }
    \subfloat{
      \centering
      \includegraphics[width=.47\textwidth]{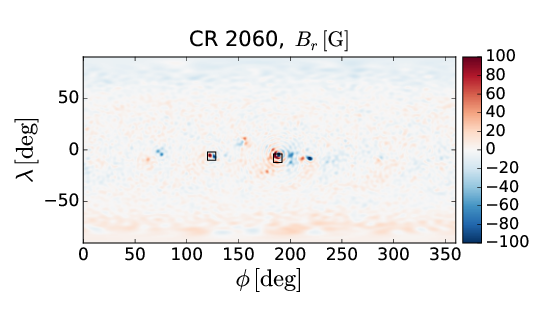}
    }\\
    \subfloat{
      \centering
      \includegraphics[width=.47\textwidth]{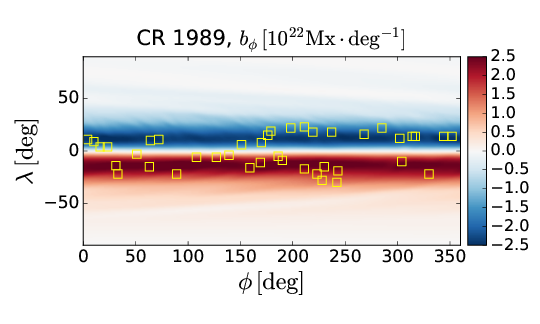}
    }
    \subfloat{
      \centering
      \includegraphics[width=.47\textwidth]{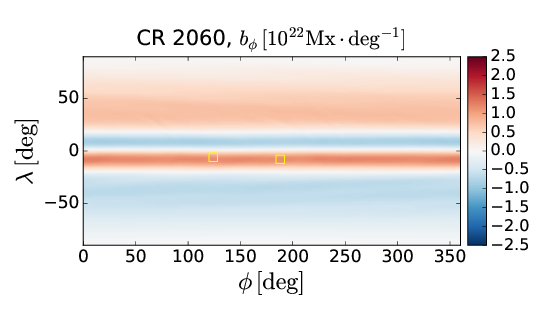}
    }\\
    \subfloat{
      \centering
      \includegraphics[width=.47\textwidth]{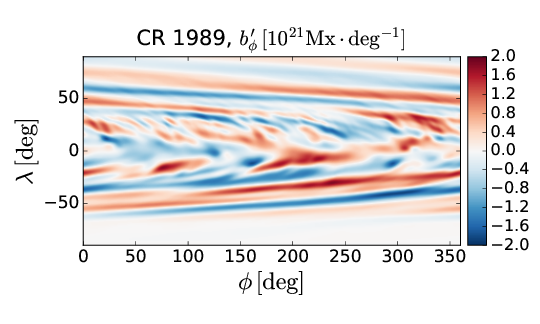}
    }
    \subfloat{
      \centering
      \includegraphics[width=.47\textwidth]{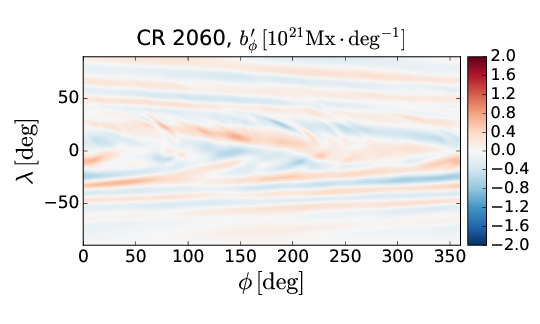}
    }\\
  \caption{Surface field ($B_r$), azimuthal flux density ($b_\phi$) and its non-axisymmetric component ($b_\phi'$), for CR 1987 (close to the middle of cycle 23) and CR 2060 (near the end of that cycle). Red and blue represent opposite polarities. The squares on the top and middle rows represent observed emergence sites, extracted from the USAF/NOAA sunspot group database.}
  \label{fig:maps}
  
\end{figure*}

The source term in Eq. \eqref{eq:sft}, which represents the emergence of flux on the solar surface, was calculated using SOLIS synoptic magnetograms. Let $B_r^n$ be the radial magnetic field corresponding to carrington rotation (CR) $n$, as given by the corresponding synoptic magnetogram. The field associated with emergences during CR $n$ was computed as
\begin{equation}
\Delta B_r(\theta,\phi,t^n) = B_r^{n}-\tilde{B}_r^{n},
\end{equation}
where $\tilde{B}_r^{n}$ is the magnetic field from the previous magnetogram evolved for one rotation using Eq. \eqref{eq:sft}. This expression is related to the surface source term $S(\theta,\phi,t)$ through
\begin{equation}
\Delta B_r(\theta,\phi,t^n) = \int_{t^{n-1}}^{t^n} S(\theta,\phi,t) \,\mathrm{d}t.
\end{equation}
In practice, we add $\Delta B_r$ to the simulated surface field every carrington rotation. The synoptic magnetograms are corrected by multiplying the positive values by a factor such that the resulting net magnetic flux on the surface is zero. Apart from describing flux emergences, the surface source term also corrects for the errors of the SFT model and errors in individual synoptic magnetograms.

The surface field $B_r$ must connect with the field in the convection zone so that $\nabla\cdot\mathbf{B}=0$ is maintained. The connectivity of the surface sources with the subsurface field is achieved through the source term in Eq. \eqref{eq:azimuthal}, which we calculated by performing a downwards potential field extrapolation of $\Delta B_r(\theta,\phi,t^n)$. Let $\Delta B_\phi(\theta,\phi,t^n)$ be the azimuthal component of the extrapolated field, and $\Delta b_\phi = \int_{R_b}^{R_\odot} \Delta B_\phi r\,\mathrm{d}r$. The source term in Eq. \eqref{eq:azimuthal}, $S_\phi(\theta,\phi,t^n)$, is related to $\Delta b_\phi(\theta,\phi,t^n)$ through
\begin{equation}
\Delta b_\phi(\theta,\phi,t^n) = \int_{t^{n-1}}^{t^n} S_\phi(\theta,\phi,t) \,\mathrm{d}t.
\end{equation}
The extrapolated $\Delta b_\phi$ is added to the simulated $b_\phi$ every rotation, at the same time $\Delta B_r$ is added to the surface field.

The potential field extrapolation of $\Delta B_r(\theta, \phi, t^n)$ is more easily done using spherical harmonics, for which it is convenient to remap the synoptic magnetograms onto a grid equally spaced in $\theta$, rather than in $\cos\theta$. The remapping prevents large errors near the poles, which arise from the poorer spatial resolution of the synoptic magnetograms at high latitudes. A discussion of this problem can be found in \cite{toth2011potential}.

The value of the magnetic field at each point of the new equiangular grid was interpolated linearly from the magnetogram (old grid). The magnetic field at the poles is not known, which makes it appropriate to perform the interpolation in Fourier space by expanding
\begin{equation}
B_r(\theta,\phi,t^n) = \sum_{m} a_m(\theta,t^n)e^{im\phi}.
\end{equation}
At the poles, regularity of $B_r$ translates into the following boundary conditions for $a_m$:
\begin{align}
& \left.\frac{\partial a_0(\theta,t^n)}{\partial\theta}\right|_{\theta=0,\pi} = 0; \label{eq:bc1}\\
& a_{m\neq0}(\theta=0,\pi;t^n) = 0. \label{eq:bc2}
\end{align}

The value of the Fourier coefficient at point of colatitude $\theta^*$ of the new grid is given by:
\begin{equation}
a_m(\theta^*,t^n)=a_m(\theta^-,t^n)+\frac{a_m(\theta^+,t^n) - a_m(\theta^-,t^n)}{\theta^+-\theta^-}(\theta^*-\theta^-),
\label{eq:newGrid}
\end{equation}
where $\theta^-$ and $\theta^+$ are the points of the old grid adjacent to $\theta^*$ in the $\theta$ direction. For the points of the new grid that are located between the north pole and the northernmost grid point of the old grid, $\theta_0^+$, Eqs. \eqref{eq:bc1} and \eqref{eq:bc2} lead to:
\begin{align}
& a_{0}(\theta^*,t^n) = a_{0}(\theta_0^+,t^n); \\
& a_{m\neq0}(\theta^*,t^n) = \frac{a_{m\neq0}(\theta_0^+,t^n)}{\theta_0^+}\theta^*.
\end{align}
For grid points between the southernmost point of the old grid, $\theta_0^-$, and the south pole, we have:
\begin{align}
& a_0(\theta^*, t^n) = a_0(\theta_0^-, t^n); \\
& a_{m\neq0}(\theta^*, t^n) = a_{m\neq0}(\theta_0^-,t^n)-\frac{a_{m\neq0}(\theta_0^-,t^n)}{\pi-\theta_0^-}(\theta^*-\theta_0^-).
\end{align}

Since the maximum angular order of the spherical harmonic analysis is limited by an anti-aliasing condition, the decomposition of $B_r$ in spherical harmonics has the effect of slightly smoothing the magnetograms (see the top row of Fig. \ref{fig:initCond}).

\begin{figure}[t]
  \centering
  \resizebox{\hsize}{!}{\includegraphics[width=\textwidth]{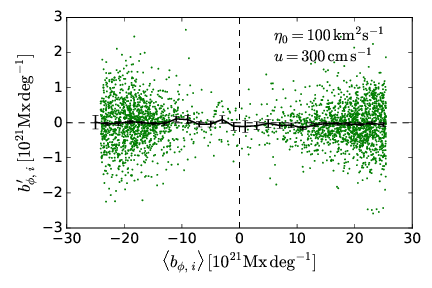}}
  \caption{Non-axisymmetric component of the azimuthal flux underlying each emergence in cycle 23 ($b_{\phi,\,i}'$) versus its axisymmetric component ($\langle b_{\phi,\,i} \rangle$). The emergences are represented by green points. The black line represents the average of $b_{\phi,\,i}'$ over all emergences inside bins of $2\cdot 10^{21}\,\mathrm{Mx\,\deg^{-1}}$ width. The error bars denote the standard error of the mean.}
  \label{fig:nonax1}
\end{figure}

\begin{figure}
  \centering
  \resizebox{\hsize}{!}{\includegraphics[width=\textwidth]{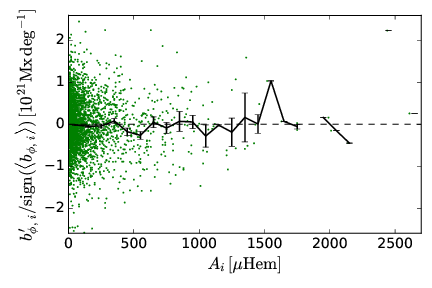}}
  \caption{Excess azimuthal flux density above the azimuthal mean underlying each emergence in cycle 23 ($b_{\phi,\,i}'/\mathrm{sign}(\langle b_{\phi,\,i} \rangle)$) versus the area of the emergence ($A_i$). The emergences are represented by green points. The black line represents the average of $b_{\phi,\,i}'/\mathrm{sign}(\langle b_{\phi,\,i} \rangle)$ over all emergences inside area bins of width $100\,\mathrm{\mu Hem}$. The error bars denote the standard error of the mean.}
  \label{fig:nonax2}
\end{figure}

\subsection{Setup and calibration}\label{sec:calib}

The initial condition of the simulations was computed as a potential field extrapolation of the first magnetogram of the series. This was taken at CR 1625, which corresponds to the end of cycle 20 in March, 1975. Figure \ref{fig:initCond} shows the raw magnetogram, the remapped version used to build the surface field source, and the extrapolated $b_\phi$ and $b_\theta$. The sign of $b_\phi$ indicates the direction of the integrated azimuthal field. In our chosen coordinate system, a positive value of $b_\phi$ corresponds to azimuthal field pointing in the sense of rotation of the Sun. The sign of $b_\theta$ is predominantly negative, which reflects the sign of the axial dipole at this cycle minimum. The irregularities of $b_\theta$ near the north pole are probably related to the noise of the magnetogram, and diffuse very quickly once the simulation starts.

To calibrate the model, we required that the simulated azimuthal flux lie radially below the active regions observed on the Sun during cycle 22. We ran simulations with different values of the free parameters for cycles 21, 22 and 23. The strength of the return flow, $u$, was varied between $1$ and $6\,\mathrm{m\,s^{-1}}$, and the diffusivity $\eta_0$ was varied between $25$ and $600\,\mathrm{km^2 s^{-1}}$. The simulations were let run for cycle 21 for initialization. The active region data (latitude, longitude and area) was extracted from the USAF/NOAA sunspot group database. We evaluate the azimuthal flux density underlying every emergence location, $b_{\phi,\,i}=b_\phi(\lambda_i, \phi_i)$, as close in time as possible, but always prior, to the time of observed maximum area of the active region (here the index $i$ runs over the emergences). In the case of backside emergences, there can be a significant delay between these two times (as large as half a rotation). We note, however, that the change in azimuthal flux density on timescales shorter than one rotation is, in most cases, small as the non-axisymmetric component represents less than $3\%$ of the total azimuthal flux density on average.

Figure \ref{fig:calibration} shows a few examples of the distribution of emergences according to their underlying azimuthal flux density for various combinations of parameters. Different values of $\eta_0$ give rise to different global magnitudes of the azimuthal flux density so, for easier comparison, we normalized $b_{\phi,\,i}$ to its maximum value in each simulation. The emergences were grouped in bins of width $4\cdot10^{-2}$ to reduce noise. The two peaks in the distribution of the active regions reflect the equatorial antisymmetry of the azimuthal flux. In cycle 22, the underlying azimuthal flux is mainly positive in the northern hemisphere and negative in the southern hemisphere. The case with $u_0=3\,\mathrm{m\,s^{-1}}$ and $\eta_0=100\,\mathrm{km\,s^{-2}}$ (black line) corresponds best to the requirement that the simulated azimuthal flux be preferentially located underneath the emergences. In the other three cases, the two-peak structure is not so conspicuous, and there are more emergences where there is little or no simulated azimuthal flux.

To find the parameter combinations that yield two well-separated peaks, we considered the quantity
\begin{equation}
\xi = \frac{\sum_c N_c |b_{\phi,\,c}|}{\sum_c N_c / |b_{\phi,\,c}|},
\end{equation}
where $c$ runs through the $b_{\phi,\,i}$ bins and $b_{\phi,\,c}$ is the mid-point of each bin.
The value of $\xi$ will be bigger for the simulations where the emergences occur farther away from the places where $b_\phi \sim 0$. Figure \ref{fig:calibration2} shows the value of $\xi$ for all test runs carried out. We find that the combination $u = 3\,\mathrm{m\,s^{-1}}$ and $\eta_0 = 100\,\mathrm{km^2s^{-1}}$ maximizes $\xi$. These parameters are close to the range found by \cite{cameron2016update} for the operation of the solar dynamo in an updated Leighton model. We proceed to the analysis of the data from cycle 23 using the azimuthal flux density maps generated in the simulation using the above parameter values.

\section{Results for cycle 23}\label{sec:results}

\subsection{Angular distribution and evolution of azimuthal flux}

Our analysis, which integrates the azimuthal field in the radial direction, allows us to infer the latitudinal and longitudinal structure of the sub-surface field from the observed surface field and large scale flows.

The top row of Fig. \ref{fig:maps} shows the observed surface field near the activity maximum of Cycle 23 (CR 1987) and towards the end of that cycle (CR 2060). Squares indicate the emergence sites from the USAF/NOAA sunspot record. Some of the emergences do not seem to correspond to strong concentrations of magnetic field in the magnetograms, and some features in the magnetograms do not have a counterpart in the active region record. A possible cause for the mismatch in the first case could be the loss of information in the low-resolution magnetograms. In the second case, one possibility is that small sunspot groups that emerged on the far side of the Sun lacked spots when the region rotated onto the visible side. In this case, the flux content of the active regions is still present in the synoptic magnetogram, and therefore included in the source term.

The middle row of Fig. \ref{fig:maps} shows the inferred maps of azimuthal flux density. The magnetic activity sits mainly on top of the azimuthal flux system.  The azimuthal flux corresponding to CR 1987 presents a structure that is strongly axisymmetric and antisymmetric about the equator. The strongest concentration of azimuthal flux occurs at $\sim 15^\circ$ of latitude in both hemispheres. At CR 2060, most of the azimuthal flux has diffused and cancelled across the equator, and a new azimuthal flux system of opposite polarity, corresponding to the new cycle, has begun to develop at higher latitudes from the winding-up of the reversed poloidal field.

The bottom row of Fig. \ref{fig:maps} shows the non-axisymmetric part of the integrated azimuthal field, calculated as $b_\phi' = b_\phi-\langle b_\phi\rangle$, where $\langle b_\phi\rangle$ is the azimuthal mean of $b_\phi$. The magnitude of $b_\phi'$ represents, on average, $\sim 3\%$ of the total azimuthal flux density. This non-axisymmetric structure arises from the emergence process (which is the only non-axisymmetric ingredient of our model), and tends to diffuse away towards the end of the cycle, when the number of emergences is smaller.

\subsection{Impact of the non-axisymmetric structure on the emergence process}

In order to investigate whether the non-axisymmetric structure of the azimuthal flux influences the emergence process, we consider the deviation of the azimuthal flux density underlying each active region from the azimuthally averaged azimuthal flux density at the latitude of emergence, $\langle b_{\phi,\,i} \rangle$. The result is shown in Fig. \ref{fig:nonax1}. As seen in Sect. \ref{sec:calib}, the bipolar distribution of events reflects the strong antisymmetry of the azimuthal field about the equator. The active regions for which $\langle b_{\phi,\,i} \rangle<0$ are mainly located in the north hemisphere, while those with $\langle b_{\phi,\,i} \rangle>0$ correspond to the south hemisphere.

\begin{figure}[t]
  \centering
  \resizebox{\hsize}{!}{\includegraphics[width=\textwidth]{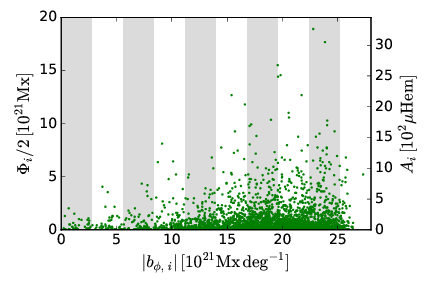}}
  \caption{Emerged magnetic flux ($\Phi_i$) versus underlying azimuthal flux density for the active regions recorded over cycle 23. Each green point represents an emergence. The stripes in the background indicate azimuthal flux density ranges. The area of the emergences is represented on the right-hand-side vertical axis.}
  \label{fig:areas1}
\end{figure}

\begin{figure}
  \centering
  \resizebox{\hsize}{!}{\includegraphics[width=\textwidth]{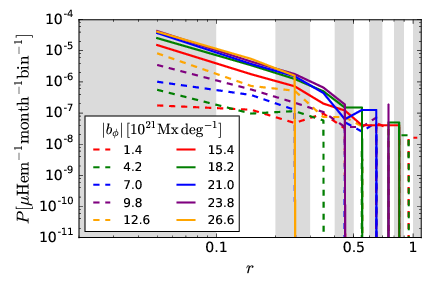}}
  \caption{Probability of emergence per unit time, per unit area and per flux ratio bin ($P$) as a function of the ratio between the emerged flux and the azimuthal flux underlying the emergence site ($r$). The different colours correspond to the ranges of azimuthal flux density represented in Fig. \ref{fig:areas1}. The numbers in the legend refer to the mid points of the azimuthal flux ranges. The stripes in the background indicate flux ratio bins of width $0.1$.}
  \label{fig:areas2}
\end{figure}

\begin{figure}
  \centering
  \resizebox{\hsize}{!}{\includegraphics[width=\textwidth]{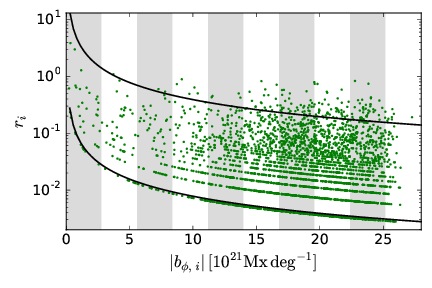}}
  \caption{Ratio of the emerged flux to the underlying azimuthal flux ($r_i$) versus the azimuthal flux density beneath the emergence location for the active regions recorded during cycle 23. Each green point represents an emergence. The stripes in the background indicate azimuthal flux ranges. The two continuous curves separate ephemeral regions (below the lower curve), medium-sized regions (between the two curves) and large active regions (above the upper curve) by their flux content, according to Table 5.1 of \cite{schrijver_activity}. These fluxes have been converted from fluxes as given by Kitt Peak magnetograms to fluxes as given by SOLIS magnetograms by using the cross-calibration constants in \cite{munoz2015areas}.}
  \label{fig:areas3}
\end{figure}

An influence of the non-axisymmetric structure of the azimuthal flux on the triggering of the emergence process would lead to a non-zero average value of $b_{\phi,\,i}'$ in each hemisphere. For example, if emergences at a given latitude tended to occur at longitudes where the azimuthal flux density is above the azimuthal mean, averaging $b_{\phi,\,i}'$ over all the emergences for which $\langle b_{\phi,\,i}\rangle>0$ would yield a positive value. In the other hemisphere, the average $b_{\phi,\,i}'$ would be smaller than zero. Computing these averages yields
\begin{align*}
\bar{b}_{\phi,\,i}' (\langle b_{\phi,\,i}\rangle<0) &= (0.1 \pm 1.7)\,\mathrm{\cdot 10^{19}Mx\,deg^{-1}};\\
\bar{b}_{\phi,\,i}' (\langle b_{\phi,\,i}\rangle>0) &= (-4.5 \pm 1.4)\,\mathrm{\cdot 10^{19}Mx\,deg^{-1}}.
\end{align*}
The average value of $b_{\phi,\,i}'$ in each hemisphere results very close to zero, in relative terms. Therefore, we do not find a significant correlation between the location of the emergence events and the departures from axisymmetry of the sub-surface azimuthal flux.

Next, we consider the possible influence of the non-axisymmetric structure on the active region areas. Figure \ref{fig:nonax2} shows the excess of azimuthal flux density above the azimuthal average beneath each active region, $b_{\phi,\,i}'/{\rm sign}(\langle b_{\phi,\,i} \rangle)$, versus the active region area, $A_i$. Again, there is no significant deviation from zero, which suggests that the inferred non-axisymmetric structure of the sub-surface azimuthal flux is unrelated to the area of the emerged active region.

\subsection{Relationship between azimuthal and emerged flux}

Here we study the relationship between the flux contained in an active region and the azimuthal flux density underneath the emergence site. Figure \ref{fig:areas1} shows the flux of each active region in the sunspot group record at the time of maximum development, $\Phi_i$, versus the unsigned underlying azimuthal flux density, $|b_{\phi,\,i}|$. Since we used SOLIS synoptic magnetograms to feed our simulations, we want to compare with fluxes comparable to those from SOLIS. The flux contained in each active region was calculated from its sunspot group area (obtained from the USAF/NOAA sunspot database) by using the cross-calibration factors in Table 2 of \cite{munoz2015areas}. The resulting relationship is $\Phi_i \,\mathrm{[Mx]}=1.44\cdot10^{19}A_i\,\mathrm{[\mu Hem]}$. A factor $1/2$ is introduced to account for the fact that the two polarities of the active region are part of a single $\Omega$-shaped magnetic structure that crosses the solar surface twice.

Using the emergences shown in Fig. \ref{fig:areas1} we estimate the probability of emergence as a function of the ratio between the flux content of the emerged active region and the azimuthal flux available within one degree colatitude directly beneath it,
\begin{equation}
    r_i = \frac{\Phi_i}{2 \int_{1\,\mathrm{deg}} b_{\phi,\,i}\,\mathrm{d}\theta}.
\label{eq:fluxRatio}
\end{equation}
To do so, we bin the data according to the ratio $r_i$ (with bins of size $0.1$) and the underlying azimuthal flux density, $b_{\phi,\,i}$ (with bins of size $2.8\cdot10^{21}\,{\rm Mx\,deg^{-1}}$). We thus obtain the number of emergences in each $(r_i, b_{\phi,\,i})$ bin. These are converted to a probability of emergence per unit area and unit time as a function of $r$ by dividing the number of emergences in each $b_{\phi,\,i}$ bin by the cycle-averaged area of the subsurface domain covered by the corresponding azimuthal flux density range, and the duration of cycle 23. The resulting probability distributions are plotted in Fig. \ref{fig:areas2}. Each coloured line corresponds to a different azimuthal flux density range. The distributions decrease rapidly for flux ratios greater than $0.4$, suggesting that emergences whose flux comprises more than $40\%$ of the azimuthal flux available underneath the emergence site are rare events.

The probability distributions shown in Fig. \ref{fig:areas2} seem to converge as we consider stronger azimuthal flux ranges. For the upper end of azimuthal flux ranges the probability of emergence is very similar. The lower probabilities obtained for emergences with smaller underlying azimuthal fluxes are due to a detection bias. To illustrate this, we plot the flux ratio of the emergences as a function of the underlying azimuthal flux density (Fig. \ref{fig:areas3}). The two curves separate ephemeral, medium and large active regions. Active regions lying closer to the lower curve have a lifetime of days, while the lifetime of those closer to the upper curve approaches weeks. Many smaller active regions will not appear in the USAF/NOAA sunspot catalogue, either because they emerge and decay on the backside of the Sun or because they do not have enough flux to form spots or pores. 
Thus, the probability distributions corresponding to lower azimuthal density fluxes (indicated by dashed lines in Fig. \ref{fig:areas2}) are substantially affected by this detection bias. The fact that the less affected distributions (corresponding to larger amounts of underlying azimuthal flux) seem to converge suggests that the probability of emergence is a function of the ratio of the emerged flux and the azimuthal flux underlying the emergence site.

\section{Summary and conclusion}\label{sec:conclusion}

We have provided a non-axisymmetric model of the magnetic flux transport in the Sun, based on the Babcock-Leighton dynamo framework. Using synoptic magnetograms as an input, we inferred the latitudinal and longitudinal distribution of azimuthal flux (per unit colatitude) and its evolution over three cycles.

We calibrated our model by requiring that the azimuthal flux in Cycle 22 in our simulations lied mainly radially underneath the activity belts. This led to a return meridional flow (and/or latitudinal pumping) having an amplitude of $u_0=3\,\mathrm{m s^{-1}}$ and an effective diffusivity for the azimuthal field of $\eta_0=100\,\mathrm{km^2s^{-1}}$. These values are in the range found by \cite{cameron2016update} for the operation of the solar dynamo. 

The azimuthal flux system is highly axisymmetric and antisymmetric about the equator. The departures from axisymmetry represent, on average, approximately $3\%$ of the azimuthal flux at a given location. We found that the non-axisymmetric structure does not have a significant impact on the location of the emergences or their observed properties. We also found that the probability of emergence is a function of the ratio of the flux content of the emerged active region and the underlying azimuthal flux.

\section*{Aknowledgements}

D.M.B. acknowledges postgraduate fellowship of the International Max Planck Research School on Physical Processes in the Solar System  and  Beyond.

This work utilizes SOLIS data obtained by the NSO Integrated Synoptic Program (NISP), managed by the National Solar Observatory, which is operated by the Association of Universities for Research in Astronomy (AURA), Inc. under a cooperative agreement with the National Science Foundation.

This work was carried out in the context of Deutsche Forschungsgemeinschaft SFB 963 "Astrophysical Flow Instabilities and Turbulence" (Project A16).

We thank Manfred Sch\"ussler for his valuable suggestions and his thorough revision of this manuscript.

\appendix

\section{The radial integral of the $\phi$ component of the induction equation}\label{ap:a}

We derive the evolution equation for the azimuthal flux per unit colatitude,
\begin{equation}
b_\phi = \int_{R_{\rm b}}^{R_\odot}B_\phi r\,\mathrm{d}r,
\label{eq:bPhiDef}
\end{equation}
under the assumptions specified in Sect. \ref{sec:model}.

The $\phi$ component of the induction equation in spherical coordinates, $(r, \theta, \phi)$, can be written as follows:
\begin{align}
    r\frac{\partial B_\phi}{\partial t} &= \frac{\partial}{\partial r} (r u_\phi B_r-r u_rB_\phi)- \frac{\partial}{\partial\theta}(u_\theta B_\phi - u_\phi B_\theta)\nonumber\\
    &-\frac{1}{\sin\theta}\frac{\partial\eta}{\partial r}\frac{\partial B_r}{\partial\phi} +\frac{\partial}{\partial r}\left[\eta\frac{\partial}{\partial r}(r B_\phi)\right]\nonumber\\
    &+ \frac{\eta}{r}\left[\frac{1}{\sin\theta}\frac{\partial}{\partial\theta}\left(\sin\theta\frac{\partial B_\phi}{\partial\theta}\right)+\frac{1}{\sin^2\theta}\frac{\partial^2B_\phi}{\partial\phi^2}\right.\nonumber\\
    &+\left.\frac{2}{\sin\theta}\frac{\partial B_r}{\partial\phi}+\frac{2\cos\theta}{\sin^2\theta}\frac{\partial B_\theta}{\partial\phi}-\frac{B_\phi}{\sin^2\theta}\right],\label{eq:indphi}
\end{align}
where we have assumed $\eta=\eta(r)$.

The first row of Eq. \eqref{eq:indphi} includes the advection and shear terms. Integrating the shear terms radially, we obtain
\begin{align}
    \left.\frac{\partial b_\phi}{\partial t}\right|_{\rm shear} &= \int_{R_{\rm b}}^{R_\odot} \frac{\partial}{\partial r}(r u_\phi B_r)\,\mathrm{d}r + \int_{R_{\rm b}}^{R_\odot} \frac{\partial}{\partial\theta}(u_\phi B_\theta)\,\mathrm{d}r\nonumber \\
    &=  \left[r u_\phi B_r \right]_{R_{\rm b}}^{R_\odot} + \frac{\partial}{\partial\theta}\int_{R_{\rm b}}^{R_\odot}u_\phi B_\theta\,\mathrm{d}r\nonumber\\
    &= R_\odot^2\Omega_{R_\odot}\sin\theta B_r|_{R_\odot}+ \frac{\partial}{\partial\theta}\int_{R_{\rm b}}^{R_\odot}r\Omega\sin\theta B_\theta\,\mathrm{d}r.\label{eq:shear1}
\end{align}
In the above equation, $B_r|_{R_{\rm b}}$ vanishes since we assume that the magnetic field does not penetrate the radiative interior. With $B_\theta=0$ in the NSSL, we can change the upper limit of integration of the integral in Eq. \eqref{eq:shear1} to $R_{\rm NSSL}$. The lower limit can be changed to the top of the tachocline, $R_{\rm T}$, since there is no poloidal field in the convection zone part of the tachocline. In the new integration domain, $\Omega$ depends only on $\theta$, so we can move it outside of the integral. This yields
\begin{align}
    \left.\frac{\partial b_\phi}{\partial t}\right|_{\rm shear} &= R_\odot^2\Omega_{R_\odot}\sin\theta B_r|_{R_\odot}\nonumber\\
    &+ \frac{\partial}{\partial\theta}\left(\Omega_{R_{\rm NSSL}}\sin\theta\int_{R_{\rm T}}^{R_{\rm NSSL}}r B_\theta\,\mathrm{d}r\right)\nonumber\\
    &= R_\odot^2\Omega_{R_\odot}\sin\theta B_r|_{R_\odot} \nonumber\\
    &+ b_\theta\sin\theta \left.\frac{\mathrm{d}\Omega}{\mathrm{d}\theta}\right|_{R_{\rm NSSL}} + \frac{\partial (b_\theta\sin\theta)}{\partial\theta}\Omega_{R_{\rm NSSL}},\label{eq:shear2}
\end{align}
where we have defined
\begin{equation}
b_\theta = \int_{R_{\rm T}}^{R_{\rm NSSL}}B_\theta r\,\mathrm{d}r.
\label{eq:bThetaDef}
\end{equation}

The radial integral of the radial advection term in Eq. \eqref{eq:indphi}, $\int_{R_{\rm b}}^{R_\odot}\partial (u_rB_\phi)/\partial r\,\mathrm{d}r$, vanishes since both $u_r$ and $B_\phi$ vanish at $R_{\rm b}$ and $R_\odot$. Integrating the latitudinal advection term yields
\begin{align}
    \left.\frac{\partial b_\phi}{\partial t}\right|_{\rm adv.} &= -\int_{R_{\rm b}}^{R_\odot}\frac{\partial}{\partial\theta}(u_\theta B_\phi)\,\mathrm{d}r\nonumber\\
    &= -\frac{\partial}{\partial\theta}\int_{R_{\rm b}}^{R_\odot}u_\theta B_\phi\,\mathrm{d}r\nonumber\\
    &= -\frac{\partial}{\partial\theta}\left[\frac{\bar{u}_R}{R_\odot}\int_{R_{\rm b}}^{R_\odot}r B_\phi\,\mathrm{d}r\right]\nonumber\\
    &= -\frac{1}{R_\odot}\frac{\partial(\bar{u} b_\phi)}{\partial\theta},
    \label{eq:adv1}
\end{align}
where $\bar{u}$ is a weighted average of the meridional flow over the convection zone,
\begin{equation}
\bar{u}(\theta) = R_\odot\overline{\left(\frac{u_\theta(r,\theta)}{r}\right)}.
\end{equation}

The diffusion term in Eq. \eqref{eq:indphi} reads:
\begin{align}
r\left.\frac{\partial B_\phi}{\partial t}\right|_{\rm diff.} 
    &=-\frac{1}{\sin\theta}\frac{\partial\eta}{\partial r}\frac{\partial B_r}{\partial\phi} +\frac{\partial}{\partial r}\left[\eta\frac{\partial}{\partial r}(r B_\phi)\right]\nonumber\\
    &+ \frac{\eta}{r}\left[\frac{1}{\sin\theta}\frac{\partial}{\partial\theta}\left(\sin\theta\frac{\partial B_\phi}{\partial\theta}\right)+\frac{1}{\sin^2\theta}\frac{\partial^2B_\phi}{\partial\phi^2}\right.\nonumber\\
    &+\left.\frac{2}{\sin\theta}\frac{\partial B_r}{\partial\phi}+\frac{2\cos\theta}{\sin^2\theta}\frac{\partial B_\theta}{\partial\phi}-\frac{B_\phi}{\sin^2\theta}\right].
\label{eq:diff1}
\end{align}
The integral of the radial part, $\int_{R_{\rm b}}^{R_\odot}\frac{\partial}{\partial r}\left[\eta\frac{\partial}{\partial r}(r B_\phi)\right]\,\mathrm{d}r$, vanishes since there is no diffusive flux transport across the boundaries. Following results of \cite{cameron2016diffusion}, we assume an effective diffusivity of azimuthal flux
\begin{equation}
\eta(r) = \eta_0\frac{r^2}{R_\odot^2}.
\label{ap:diffExpl}
\end{equation}
Substituting the above expression in Eq. \eqref{eq:diff1} and integrating the rest of the diffusion terms leads to:
\begin{align}
\left.\frac{\partial b_\phi}{\partial t}\right|_{\rm diff.} 
    &=\frac{\eta_0}{R_\odot^2}\left[\frac{1}{\sin\theta}\frac{\partial}{\partial\theta}\left(\sin\theta\frac{\partial b_\phi}{\partial \theta} \right)\right. \nonumber\\
     &+ \left. 2\frac{\cos\theta}{\sin^2\theta}\frac{\partial b_\theta}{\partial\phi}-\frac{b_\phi}{\sin^2\theta} + \frac{1}{\sin^2\theta}\frac{\partial^2 b_\phi}{\partial \phi^2} \right].
    \label{eq:diff2}
\end{align}

Combining Eqs. \eqref{eq:shear2}, \eqref{eq:adv1}, and \eqref{eq:diff2}, and introducing the source term that is necessary to ensure connectivity with the surface sources, we obtain the final form of the equation:
\begin{align}
    \frac{\partial b_\phi}{\partial t} & = R_\odot^2\,\Omega_{R_\odot} \sin\theta\, B_r \nonumber\\
    & + b_\theta\sin\theta \left.\frac{\mathrm{d}\Omega}{\mathrm{d}\theta}\right|_{R_{\rm NSSL}} + \frac{\partial (b_\theta\sin\theta)}{\partial\theta}\Omega_{R_{\rm NSSL}} \nonumber\\
    & + \frac{\eta_0}{R_\odot^2}\left[\frac{1}{\sin\theta}\frac{\partial}{\partial\theta}\left(\sin\theta\frac{\partial b_\phi}{\partial \theta} \right) + 2\frac{\cos\theta}{\sin^2\theta}\frac{\partial b_\theta}{\partial\phi}-\frac{b_\phi}{\sin^2\theta} \right. \nonumber\\
    & + \left. \frac{1}{\sin^2\theta}\frac{\partial^2 b_\phi}{\partial \phi^2} \right] - \frac{1}{R_\odot}\frac{\partial ( \bar{u} b_\phi)}{\partial\theta} + S_\phi(\theta,\phi,t).
    \label{eq:azimuthalRep}
\end{align}
To simplify the notation and keep consistency with Eq. \eqref{eq:sft}, we write $B_r$ instead of $B_r|_{R_\odot}$ to refer to the radial field at the surface.
\section{Calculation of $b_\theta$}\label{ap:b}

The quantity $b_\theta$ can be calculated in terms of $b_\phi$ and $B_r|_{R_\odot}$ from the solenoidality condition, $\nabla\cdot\mathbf{B}=0$. Writing the divergence operator in spherical coordinates leads to
\begin{equation}
r\frac{\partial(B_\theta\sin\theta)}{\partial\theta}=-\sin\theta\frac{\partial(r^2 B_r)}{\partial r}-r\frac{\partial B_\phi}{\partial\phi}.
\end{equation}
Integrating over the convection zone and using the definitions \eqref{eq:bPhiDef} and \eqref{eq:bThetaDef} we obtain
\begin{align}
    \frac{\partial(b_\theta\sin\theta)}{\partial\theta}&=-\sin\theta[r^2 B_r]_{R_{\rm b}}^{R_\odot}-\frac{\partial b_\phi}{\partial\phi}\nonumber\\
    &=-\sin\theta R_\odot^2 B_r|_{R_\odot}-\frac{\partial b_\phi}{\partial\phi},
\end{align}
where we have used $B_r|_{R_{\rm b}}=0$ and $B_\theta|_{R_{\rm b}<r<R_{\rm T}} = B_\theta|_{R_{\rm NSSL}<r<R_\odot} = 0$. Integrating now in $\theta$ yields
\begin{equation}
    b_\theta =-\frac{1}{\sin\theta}\left(\int_0^\theta \sin\theta' R_\odot^2 B_r \,\mathrm{d}\theta'+\int_0^\theta \frac{\partial b_\phi}{\partial\phi}\,\mathrm{d}\theta'\right),
\end{equation}
where, again, $B_r$ denotes now the surface field.

\bibliographystyle{aa}
\bibliography{/home/david/texmf/bibtex/reviews,/home/david/texmf/bibtex/papers,/home/david/texmf/bibtex/books}

\end{document}